# GANkyoku: a Generative Adversarial Network for Shakuhachi Music

**Omar Peracha**
Humtap
omar.peracha@gmail.com

**Shawn Head**
Independent
shawnheadmusic@gmail.com

## ABSTRACT

*A common approach to generating symbolic music using neural networks involves repeated sampling of an autoregressive model until the full output sequence is obtained. While such approaches have shown some promise in generating short sequences of music, this typically has not extended to cases where the final target sequence is significantly longer, for example an entire piece of music. In this work we propose a network trained in an adversarial process to generate entire pieces of solo shakuhachi music, in the form of symbolic notation. The pieces are intended to refer clearly to traditional shakuhachi music, maintaining idiomaticity and key aesthetic qualities, while also adding novel features, ultimately creating worthy additions to the contemporary shakuhachi repertoire. A key subproblem is also addressed, namely the lack of relevant training data readily available, in two steps: firstly, we introduce the PH_Shaku dataset for symbolic traditional shakuhachi music; secondly, we build on previous work using conditioning in generative adversarial networks to introduce a technique for data augmentation.*

## 1. INTRODUCTION

Recent years have seen several new proposals for generating symbolic music with deep neural networks [1, 2, 3, 4, 5, 6, 7, 8]. Many of these comprise training of a single network, typically consisting of layers of some recurrent unit such as LSTM [9], and predicting the target sequence one time step at a time. The task of music generation has proven sufficiently complex so as to limit the overall lengths of sequences whose inherent structural-temporal features are learnable by these approaches, whereby generated samples consistently demonstrate those same key features. Consequently, complete pieces of music of any moderate length have been difficult to synthesise in their entirety using neural networks, unless a lack of adherence to the musical qualities of the training data over time is acceptable to the composer utilising the network.

While some recent works have demonstrated the ability of neural networks to generate different styles of music, from Bach [1] to the Blues [2], comparatively little work has been done regarding generating music with specific instruments in mind. This is counter to a situation composers may commonly find themselves in, where they are commissioned to produce music not only with a certain stylistic expectation, but for a specific instrumentation.

In this work, we propose a deep neural network to generate entire pieces written specifically for the shakuhachi. Our aim for these pieces is that they should maintain key aesthetic qualities expected of shakuhachi music, derived from the traditional repertoire, including long term structural-temporal features. They should also be idiomatic from a performer's perspective, which includes utilising a notation system specific to shakuhachi music. We do not, however, purely want plain pastiches of classical shakuhachi music; rather, we seek for a model that, through its own (possibly nonintuitable) idiosyncratic qualities, adds novel features in a systemic way, thus having the ability to make contributions to the contemporary shakuhachi repertoire which some humans may subjectively deem to be worthwhile. These contributions should demonstrate variety among themselves, including in their lengths.

Generative Adversarial Networks have been proposed as a framework for training generative models [10], and they have shown a great capacity to generate completely original samples which clearly reflect important features of the training data; for example, [11] generated convincing images of human faces which were unseen in the training set. Furthermore, [12] leveraged conditioning the input of a GAN in order generate realistic samples of single cell RNA-seq data when few real samples of such sequences were first observed in the training set. We build on these findings to introduce a technique for data augmentation, which we then apply to the process of training our own model.[1]

## 2. RELATED WORK

Symbolic monophonic melody generation at the note level with deep recurrent neural networks has been illustrated by [3] with their well-known MelodyRNN model, while [3, 4] and [5] have generated melodies with a larger base unit. In [6], an LSTM-based GAN is used to generate melodies, representing the data as a fully continuous

[1] All code, data and samples for this project can be found at the following link: *https://github.com/omarperacha/GANkyoku*

sequence rather than a discrete sequence of distinct note categories.

MidiNet, a conditional GAN for music generation, was proposed by [7] and can generate symbolic melodies one bar at a time, conditioned by a chord progression and a priming melody. Unlike the majority of music generation approaches seen to date, MidiNet uses convolutional layers. Temporal dependencies are not sufficiently learned by the convolutional layers in this approach, and so conditioning on a previous part of the sequence is required to generate melodies of more than one bar in length.

Generating polyphonic symbolic music with longer-form structures was illustrated by [1] and [8], with the former synthesising Bach chorals and the latter outputting pop songs with relatively simplistic arrangements consisting of melody, drums and a chordal accompaniment. Few of the available output samples from either of these models are significantly longer than one minute in length, however, suggesting that they were not able to learn features which keep music in those styles engaging over longer periods of time.

Given the nature of shakuhachi notation used in this work, there is a perceptible relevance to natural language generation in this task. Char-rnn [13] demonstrates the ability to generate samples of text at the character level which strongly correlate with the training set, though it is not able to maintain clear themes or topics for more than a few words in its outputs. In [14], a GAN architecture is proposed which places the adversarial training process in a reinforcement learning framework in order to optimise it specifically for the task of modeling sequences of discrete tokens, including Chinese poetry. This seems a promising approach to sequence generation with adversarial networks, however our work favours a continuous representation of the data.

Finally, a GAN is used in [12], consisting simply of affine layers, to generate single cell RNA-seq data. Through training over a large dataset consisting of multiple classes of related data, all of which they wished to model, and conditioning both the generator model and the discriminator model on the class label, they were able to generate realistic samples of the specific classes observed in their training set, including a class for which there were comparatively very few samples present during training. We leverage their findings by first synthesising otherwise non-existent classes of data related to that which we wish to model, and similarly conditioning our network on the class label, allowing us to dramatically augment our training set and better learn features of the single class with which we are in fact concerned.

## 3. DATASET

### 3.1 Shakuhachi Music

The shakuhachi is traditionally associated with monks from the *Fuke* school of Zen Buddhism. The classical repertoire, or *honkyoku*, comprises pieces played by these monks for the purposes of seeking both alms and enlightenment. Common perception of the shakuhachi's sound and music is strongly influenced by its Zen origins, with many contemporary pieces for shakuhachi continuing to make use of its idiosyncratic qualities in ways which reflect and reference these origins.

Today, there are multiple schools of shakuhachi playing, which can differ to varying degrees in playing style, notation and repertoire. The two most popular schools of playing are *Kinko Ryu* and *Tozan Ryu*. The former has a greater focus on the traditional repertoire, which numbers at approximately 40 pieces [15]. Because of its greater focus on these 40 original *honkyoku*, we opt for *Kinko Ryu* notation as the method of symbolicating the music observed and output by our network. A further reason for this preference is that both authors of this work have previously studied shakuhachi playing from the *Kinko* lineage, with the second author considered an expert.

### 3.2 Notation

*Kinko Ryu* notation is different to Western music notation and unique to the shakuhachi. The sheet music consists of columns of predominantly Katakana characters, read from top to bottom, each of which may indicate a note, playing technique or other metadata such as specifying an octave. The columns themselves are read sequentially from right to left. See Figure 1 for an example of *Kinko Ryu* notation.

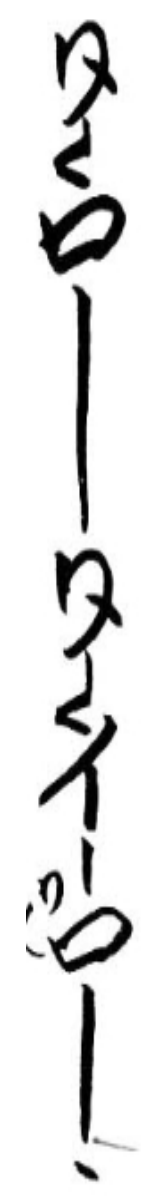

**Figure 1**. A single phrase of shakuhachi music in *Kinko* notation [16].

Certain characters refer to a specific pitch, while others may refer simply to the pitch class, with information about the octave coming from a separate symbol or from traditional convention. A key feature of *honkyoku* is that rhythmic durations are usually unspecified. The length of a vertical line separating two characters may be used as a rough guide to inform playing length, but in general rhythmic notation is far less precise than in Western notation. This contributes greatly to shakuhachi music's particular aesthetic qualities. This also provides an advan-tage to *honkyoku* in terms of their fitness to use as the basis of a generative model, because less musical information needs to be encoded into the symbolic representation, theoretically reducing the complexity of the sequence to be modeled.

Similarly, form is observed to be flexible across the *honkyoku* repertoire; while certain patterns do emerge, such as the middle third of a piece usually showing the greatest intensity, or certain gestural shapes tending to recur throughout a piece, there are no clear-cut rules that must be followed. The compositional process behind many pieces would appear to be relatively intuitive, with a greater concern for dependencies between local events than long-term structural dependencies. In some senses this may further reduce the complexity of the sequence to be modeled, however it also serves to reduce the correlation between different samples in the training set, potentially causing any long-term structural qualities which do exist to be less apparent.

### 3.3 Dataset and Representation

#### *3.3.1 PH_Shaku Dataset*

Given that our goal was to capture qualities from the *honkyoku* repertoire and add novel features, it follows that these *honkyoku* should feature in the set of data observed by the model during training. Unfortunately, these pieces are not available in any representation other than the sheet music described above. Because the notation has idiosyncratic qualities, is typically handwritten, and relatively few pieces exist in the first instance, automated transcription of the *Kinko Ryu* set of *honkoyu* pieces is beyond the scope of this work. We therefore went through the process of transcribing some of these pieces manually.

These transcriptions are made available publicly as the PH_Shaku dataset, and consist of CSV files where each file contains a tokenised representation of a single piece, made entirely from ASCII characters. Each unique symbol observed in the sheet music is given its own unique token, usually consisting of a romanised version of the character's name, or that of the playing technique it represents. We also add start and end tokens, bringing the total number of distinct tokens to 45.

#### *3.3.2 Representation*

As described in [14], GANs can show limitations when trying to generate sequences of discrete tokens. Secondly, when calculating the cost of an incorrect prediction in a classification task, typical loss functions, such as categorical cross entropy, treat all incorrect predictions with equal weighting; that is to say that there is no concept of some predictions being more wrong than others. In many music generation tasks, this may not provide the optimal method to train a model. For example, an incorrectly predicted note which is also in the wrong key could in many contexts be a worse prediction than an incorrect note which belongs in the underlying chord. It is not an arbitrary task to decide how such a hierarchical penalty system should be constructed, as this would clearly depend to a great extent on several contextual and musical factors. While future work may consider creating a model to aid in this task, or introducing new methods of calculating the loss for music-specific problems, we approached this problem with a naive method.

We represent our data as continuous, as a preprocessing step. First, we manually assign each of our tokens an integer label, from 0 to 44, with the intention that those tokens whose labels are close to one another are more closely contextually related. The start token is 0, the end token is 44, and the token which signifies breathing points, a key structural marker, is given the central value of 22. These values are then normalised to have a mean of 0 and a standard deviation of 1, meaning the aforementioned tokens are remapped to -1, 1 and 0 respectively. Thus our symbolic music data is represented as values between -1 and 1, with elements of the final output sequence being rounded to the nearest value which corresponds to a token mapping, before being converted back into the human-readable token for evaluation.

#### *3.3.3 Dataset Augmentation*

The process of manually converting handwritten handwritten *honkyoku* into a tokenised dataset was very tedious. Furthermore, there are only a maximum of 40 training samples which could be obtained in this method. This would typically considered a very small dataset and seemed likely to need augmentation. We therefore stopped tokenising *honkyoku* once we had converted ten such pieces, which demonstrated variety and ranged in length from 67 to 576 elements. Instead, we turned our focus to synthesising more data.

In [12], conditioning of a GAN was used in order to generate realistic samples of a class of data of which there were few examples in the training data. This is deemed to be because the model was able to learn those features it shared with the remaining classes from the examples of those other classes, which were far more abundant in the training data.

Our aim was not to generate realistic synthetic examples of the training data, but to introduce new qualities to samples which also perceivably correlate to the *honkyoku* repertoire. Training a model to try approximate the *honkyoku* closely did seem a reasonable approach, however, because of the small data set; we could anticipate that the model might add unique and interesting musical features precisely corresponding to the particular way in which it fails to grasp some of the *honkyoku*'s features. Should the model generate samples too close to the classical repertoire, we could introduce temperature during sampling, or simply use a different set of weights.

The problem remained that we had a small amount of *honkyoku* pieces available to train on. We therefore leveraged the findings of [12] to create a framework in which we could successfully train our model. Making use of the fact our data was represented as continuous real values, we created synthetic data by adding a controllable amount of noise to our real-world data. Firstly, we define a variable N for the noise factor that determines the extent of the noise being added to the original sample, where $0 <= N <= 1$. Then, for each value in the original sequence besides the start and end tokens, we multiply by a number chosen randomly from the range (1-N, 1+N). Finally, we return the hyperbolic tangent of the obtained value, ensuring all values in the new sequence remain between -1 and 1. Thus we have a new sequence with a controllable amount of dissimilarity from the original real-world example. All sequences are end-padded with end tokens up to the maximum sequence length of 576.

A key quality of this function is that an input 0 remains as 0 when output. In our data mapping, 0 equates to the symbol which marks a breath or phrase end. The positioning of these symbols is deemed by us to be a key contributing factor to creating the appropriate feel in *honkyoku*. From a more practical perspective, it is important that the performer is given sufficient opportunity to breathe. Therefore we are able to synthesise data that varies from that which we are trying to model most closely, but shares key structural characteristics.

We now classify our total dataset, *honkyoku* and synthetic samples, according to the degree of dissimilarity from a real sample, as dictated by the value of N. A class label of 0 is given to an instance of an actual *honkyoku* piece, 1 for 0 < N < = (8/25), 2 for (8/25) <= N < (17/25) and 3 for any other value of N, creating a total of four classes. The class label is provided to both the generator and discriminator models as inputs during training. In this way we could recreate the circumstances seen in [12] to improve our likelihood of generating samples demonstrating the intended characteristics.

## 4. MODEL

Our model consists of an LSTM-based generator, G, and a convolutional discriminator model, D, trained in adversarial process to minimise the Wasserstein distance, which has been shown to improve the often unstable process of training a GAN [17, 18]. Following convention for so-called Wasserstein GANs, D is trained five times for every single epoch that the G is trained, and RMSProp, with the a learning rate set to 0.00005, is used as the optimiser.

G's input is a noise vector of length 128, and a one-hot encoded vector representing the label of the class trying to be generated. The two inputs are concatenated end to end and fed into a dense layer with 576 units, followed by two 1024-unit LSTM layers with dropout applied to each. The output tensor is then fed into another 576-unit dense layer with a tanh activation function, which in turn outputs the predicted sample.

D's input is either a sample generated by G or a sample from the training set, and a one-hot encoded vector representation of the intended or actual class label. The input sample is fed through four 1D convolutional layers with an increasing number of filters, a consistent filter size of 2, and no zero-padding. A leaky ReLU is used as the activation function of these layers. All these layers use dropout, and all but the first undergo batch normalisation. The final output of the last convolutional layer is end-concatenated with the condition vector, before being input to a single unit dense layer.

Training was undertaken on a GPU with a batch size of 100, across 17,701 batches, taking approximately two days.

## 5. EVALUATION

Our outset was specifically to create novel sounding and engaging music, which creates a challenging environment for evaluation. Firstly, metrics such as prediction accuracy or positive correlation to training data are not directly relevant, because we are intentionally trying to create differences that would score lower in these analyses, and there is otherwise no clear guidance on how to interpret these scores in order to evaluate the performance in the given task. In terms of pure training metrics, GANs are notoriously difficult to assess because improvements in the loss do not necessarily correlate with improvements in the output samples. While minimising the Wasserstein distance reduces the opacity of GAN loss metrics to an extent, one still has to inspect the samples to truly gauge performance. Secondly, determining the success of a piece of music can be highly subjective in many respects. Consequently, the most effective method of evaluation is possibly for humans to inspect the audio and symbolic text samples contained among the files in the code repository. The overall evaluation process would in this case be much like how a composer would first evaluate their work themselves, followed by evaluation an audience and critics upon public release.

While a lack of an appropriate scientific evaluation method may be deemed unsatisfactory, one fact remains true of the model's outputs which from a practical perspective is arguably the most important measure of their success; they are regularly performed in a concert environment. At the time of writing, 30 samples have been generated by our model using the optimal set of weights obtained during training. Three of these have been played by professionals in concert programmes consisting of both traditional and contemporary shakuhachi works. However, a variance in quality across the set of all generated outputs has been observed.

Several positives have been discerned by examining the generated samples, and also some room for future improvement. Among the positives is that variety in output samples is clearly demonstrated. Output pieces vary greatly in length, with samples ranging from 78 tokens to 568 observed so far. They also demonstrate differences in features such as their phrase lengths, the consistency of phrase lengths over time, types of gesture appearing within the phrases and consistency of gestural shapes across time. Crucially, pieces can maintain a perceivable musical trajectory with relative consistency right until the end of the piece. That said, there can still be occasions in outputs when this trajectory is lost and the piece seems to wander aimlessly, though often it will find its way back.

Other than some irregularities in development of material over a long term structure, there are some instances of unideal tendencies on a more local scale; there may be instances of two consecutive breath tokens, for example, which could stem from the technique for augmentation technically not preventing this from occurring in the syn-thesised data. Secondly, there can be occurrences of several symbols tokenised as *meri* occurring in a row. This is likely a combined result of *meri* being a very common symbol in the training data, and its relative positioning in the naive mapping to continuous values of the tokens being unideal. Finally, some outputs may immediately start off on a path that would perplex a player, usually coming to an abrupt end. This may be solvable by a more effective method of conditioning, such as in [12].

## 6. CONCLUSIONS

We have demonstrated that a generative adversarial network can be used to model entire pieces for shakuhachi, in a symbolic form. We have shown that these pieces can demonstrate variety, musical sense over long periods of time and idiomaticity. We have also shown that these pieces can feature some novelty and creativity, without losing reference to aesthetic expectations set by historical associations.

We introduced a technique to augment datasets of continuous sequences, and improve the ability to model a dataset with a small number of such samples by conditioning the model to see the real-world samples as belonging to one class, and the synthesised samples as belonging to other classes. This may be helpful in the cases where a composer wants to model music in styles which do not have large datasets available.

In doing so, we have explored the viability of deep neural networks today for playing a significant role in the composition of new pieces with musically forward-thinking intent. We have demonstrated how this is possible, but there remains room to improve the model's ability to meet expectations, for example by better representing the musical data or creating cost functions which are optimiised for this purpose.